\documentclass[aps,prb,twocolumn,showpacs]{revtex4-1}

\usepackage{graphicx}
\DeclareGraphicsExtensions{.png,.jpg,.eps}
\usepackage[utf8]{inputenc}

\usepackage{xcolor}
\usepackage{amsmath}
\usepackage{blindtext, rotating}
\usepackage{braket}
\usepackage{lipsum}
\usepackage{mathtools}

\begin{document}

\title{One-to-one correspondence between thermal structure factors and coupling constants of general bilinear Hamiltonians}

\author{Bruno Murta}
\affiliation{QuantaLab, International Iberian Nanotechnology Laboratory (INL), 4715-330 Braga, Portugal}
\affiliation{Departamento de F\'{i}sica, Universidade do Minho, 4710-057 Braga, Portugal}

\author{ J. Fern\'andez-Rossier\footnote{On leave from Departamento de F\'isica Aplicada, Universidad de Alicante,  Spain
}, }
\email{joaquin.fernandez-rossier@inl.int}
\affiliation{
QuantaLab, International Iberian Nanotechnology Laboratory (INL), 4715-330 Braga, Portugal
}

\date{\today}

\begin{abstract}
A theorem that establishes a one-to-one relation between zero-temperature static spin-spin correlators and coupling constants for a general class of quantum spin Hamiltonians bilinear in the spin operators has been recently established by J. Quintanilla, using an argument in the spirit of the Hohenberg-Kohn theorem in density functional theory. Quintanilla's  theorem gives a firm theoretical foundation to quantum spin Hamiltonian learning using spin structure factors as input data. Here we extend the validity of the theorem in two directions. First, following the same approach as Mermin, the proof is extended to the case of finite-temperature spin structure factors,
thus ensuring that the application of this theorem to experimental data is sound.
Second, we note that this theorem applies to all types of Hamiltonians expressed as sums of bilinear operators, so that it can also relate the density-density correlators to the Coulomb matrix elements for interacting electrons in the lowest Landau level.
\end{abstract}

\maketitle

Understanding the  wonders and complexities of the microscopic world requires tackling the notoriously hard quantum many-body problem. Given the ubiquity of approximate methods in the state-of-the-art research on quantum many-body phenomena, the existence of general theorems\cite{hohenberg64,mermin65,mermin66,Lieb1989} is essential to set such approximations on firm theoretical ground. 

The present Letter follows a recent work by J. Quintanilla \cite{quintanilla22}, which establishes a theorem valid for a general class of bilinear quantum spin Hamiltonians, 
 \begin{equation}
\hat{H}=\sum_{i,j,\alpha,\beta}  J_{i,j}^{\alpha,\beta} \hat{S}_i^{\alpha}\hat{S}_j^{\beta},
\label{H}
\end{equation}
where $\hat{S}_i^{\alpha}$ is the $\alpha=x,y,z$ component of a spin operator acting on site $i$ in an arbitrary lattice, and $J_{i,j}^{\alpha,\beta}$ are the spin coupling constants. The general Hamiltonian stated in eq. (\ref{H}) encompasses most physically relevant types of interactions, notably Heisenberg\cite{Heisenberg28}, dipolar, Ising\cite{Ising25}, Dzyaloshinskii–Moriya\cite{Dzyaloshinsky58,Moriya60} and Kitaev\cite{Kitaev06} interactions. As a result, a wide class of canonical quantum spin models (e.g., Ising\cite{Ising25}, Heisenberg\cite{Heisenberg28}, XXZ\cite{Orbach58}, Majumdar–Ghosh\cite{Majumdar1969, Majumdar1969a}, Shastry-Sutherland\cite{Shastry81}, Haldane\cite{Haldane1988}, Kitaev\cite{Kitaev06}) are particular cases of eq. (\ref{H}). It should be noted, however, that this class of Hamiltonians does not cover  important  quantum spin models such as the toric code\cite{Kitaev03} or the bilinear-biquadratic Heisenberg model (including the AKLT model\cite{Affleck1987}), which is known to describe some physical systems\cite{Mishra2021}.

The theorem proven by J. Quintanilla\cite{quintanilla22} for this class of bilinear spin Hamiltonians asserts that there exists a one-to-one correspondence between the exchange constants $J_{i,j}^{\alpha,\beta}$ and the zero-temperature correlators
\begin{equation}
\rho_{i,j}^{\alpha,\beta} (T=0)  = \langle \Phi_0| \hat{S}_i^{\alpha}\hat{S}_j^{\beta} |\Phi_0\rangle
\label{correlator_T0}
\end{equation}
for a physical system represented by the wave function $\ket{\Phi_0}$, which corresponds to the non-degenerate\footnote{The proof of Quintanilla's theorem assumes the ground states of the two different bilinear spin Hamiltonians are both non-degenerate, but a follow-up discussion relaxes this assumption.} ground state of a Hamiltonian of the form given in eq. (\ref{H}). The proof of Quintanilla's theorem runs parallel to that of the Hohenberg-Kohn theorem\cite{hohenberg64} for density functional theory.  Interestingly, Mermin generalized\cite{mermin65} the Hohenberg-Kohn theorem to finite temperature, which motivates us to look for a finite-temperature generalization of Quintanilla's theorem as well.

Before proceeding to the extension of Quintanilla's theorem to the case of finite temperature, we first introduce some relevant concepts and notation. The thermal spin correlators at temperature $T$ are defined as
\begin{equation}
\rho_{i,j}^{\alpha,\beta} (\hat{W}) \coloneqq \textrm{Tr}\left( \hat{W} \hat{S}_i^{\alpha}\hat{S}_j^{\beta}\right),
\label{correlator_T_abs}
\end{equation}
where $\hat{W}=\sum_n \frac{e^{-\beta E_n}}{Z} |\Phi_n\rangle\langle \Phi_n|$ is the density operator, $Z=\sum_n e^{-\beta E_n}$ is the partition function, $\{E_n\}$ and $\{|\Phi_n\rangle\}$ are the eigenenergies and eigenstates of a Hamiltonian $\hat{H}$ of the form given in eq. (\ref{H}), and $\beta^{-1} = k_B T$. Expanding eq. (\ref{correlator_T_abs}) in the Hamiltonian eigenbasis gives
\begin{equation}
\rho_{i,j}^{\alpha,\beta} (\hat{W}) = \frac{1}{Z}
\sum_n e^{-\beta E_n}\langle \Phi_n|
 \hat{S}_i^{\alpha} \hat{S}_j^{\beta}|\Phi_n\rangle. 
\label{correlator_T_eigbasis}
\end{equation}
Setting $T = 0$, or $\beta \to \infty$, in eq. (\ref{correlator_T_eigbasis}) results in eq. (\ref{correlator_T0}), as expected. We note that spin correlators can be measured experimentally using neutron diffraction\cite{shull1949,marshall1968}.

 In the following we show that the mapping between the coupling constants $J_{i,j}^{\alpha,\beta}$ and the {\em finite-temperature} correlators (cf. eq. (\ref{correlator_T_eigbasis})) is bijective. We restrict ourselves to finite temperatures, since for $T \to \infty$ all but local $\langle \hat{S}_i^{z} \hat{S}_i^{z}
\rangle$ correlators vanish (cf. Appendix A).  

The proof follows in a similar vein to the zero-temperature one by Quintanilla\cite{quintanilla22}, but the Rayleigh-Ritz variational principle is replaced by the Gibbs-Bogoliubov inequality\cite{Gibbs10, mermin65} for the Helmholtz free energy.  Let a system described by a Hamiltonian $\hat{H}$ be in contact with a thermal bath at temperature $T$. The Helmholtz free energy of such system is
\begin{equation}
F(\hat{W})=-k_B T \ln Z = \langle \hat{H} \rangle_{\hat{W}} - T S[\hat{W}],
\label{F}
\end{equation}
where $\langle \hat{O} \rangle_{\hat{W}}  = \textrm{Tr} \left( \hat{W} \hat{O} \right) = \frac{1}{Z} \sum_n e^{-\beta E_n}\langle \Phi_n|\hat{O}|\Phi_n\rangle$ is the expectation value of some operator $\hat{O}$ at finite temperature and $S[\hat{W}] = -k_B \textrm{Tr} \left( \hat{W} \ln \hat{W} \right)$ is the von Neumann entropy. The Gibbs-Bogoliubov inequality sets an upper bound on the Helmholtz free energy,
\begin{equation}
F(\hat{W})\leq \langle \hat{H}\rangle_{\hat{W}^{'}} - T S[\hat{W}^{'}],
\label{Bogo}
\end{equation}
for any positive semidefinite operator $\hat{W}^{'}$ of appropriate dimensionality. The equality in eq. (\ref{Bogo}) only occurs either when $\hat{W} = \hat{W}^{'}$ or $T \to \infty$. A proof of the Gibbs-Bogoliubov inequality can be found in Mermin\cite{mermin65}.

The proof of Quintanilla's theorem at finite temperature proceeds by {\it redutio ad absurdum}. We consider two different Hamiltonians of the form given in eq. (\ref{H}), $\hat{H}$ and $\hat{H}^{'}$. Their corresponding coupling constants,
$J_{ij}^{\alpha,\beta}$ and $J_{ij}^{'\alpha,\beta}$, cannot therefore be all equal in pairs. Since the coupling constants determine the energies and eigenstates, they determine $\hat{W}$ and $\hat{W}^{'}$ as well, the equilibrium density operators for $\hat{H}$ and $\hat{H}^{'}$, respectively. We then assume that both $\hat{W}$ and $\hat{W}^{'}$ are associated with the same finite-temperature spin-spin correlators,
\begin{equation}
\rho_{i,j}^{\alpha,\beta}[\hat{W}]=\rho_{i,j}^{\alpha,\beta}[\hat{W}^{'}],
\label{false}
\end{equation}
for all $i,j,\alpha,\beta$. We can use the Gibbs-Bogoliubov inequality to write the following expression for the Helmoltz free energy of the unprimed system:
\begin{widetext}
\begin{align}
F(\hat{W}) & = \sum_{i,j,\alpha,\beta} J_{ij}^{\alpha,\beta} \rho_{i,j}^{\alpha,\beta}[\hat{W}] - T S[\hat{W}]
\leq  \sum_{i,j,\alpha,\beta} J_{ij}^{\alpha,\beta} \rho_{i,j}^{\alpha,\beta}[\hat{W}^{'}] - T S[\hat{W}^{'}]=\nonumber\\
 & =
 \sum_{i,j,\alpha,\beta} \left(J_{ij}^{\alpha,\beta}- J_{ij}^{'\alpha,\beta}\right)\rho_{i,j}^{\alpha,\beta}[\hat{W}^{'}] + F(\hat{W}^{'}).
  \label{inequality_unprimed}
 \end{align}
We can now exchange the roles of $\hat{W}$ and $\hat{W}^{'}$ to obtain an identical expression for the primed system:
\begin{equation}
F[\hat{W}^{'}]\leq
 \sum_{i,j,\alpha,\beta} \left(J_{ij}^{'\alpha,\beta}- J_{ij}^{\alpha,\beta}\right)\rho_{i,j}^{\alpha,\beta}[\hat{W}] + F(\hat{W}).
 \label{inequality_primed}
\end{equation}
Summing eqs. (\ref{inequality_unprimed}) and (\ref{inequality_primed}) yields:
\begin{eqnarray}
F[\hat{W}]+F[\hat{W}^{'}]\leq \sum_{i,j,\alpha,\beta} \left(J_{ij}^{'\alpha,\beta}- J_{ij}^{\alpha,\beta}\right)
\left(\rho_{i,j}^{\alpha,\beta}[\hat{W}]-\rho_{i,j}^{\alpha,\beta}[\hat{W}^{'}]\right)+ F[\hat{W}]+F[\hat{W}^{'}].
\label{eq_final}
\end{eqnarray}
\end{widetext}
Using eq. (\ref{false}) turns eq. (\ref{eq_final}) into: $F[\hat{W}]+F[\hat{W}^{'}]\leq F[\hat{W}]+F[\hat{W}^{'}]$. The equality holds only in the two trivial limits of infinite temperature or $\hat{H} = \hat{H}^{'}$. For finite temperature and $\hat{H} \neq \hat{H}'$, we can replace the symbol $\leq$ by a strict inequality, thus arriving at a contradiction: $F[\hat{W}]+F[\hat{W}^{'}]< F[\hat{W}]+F[\hat{W}^{'}]$.  It follows, then, that the initial assumption stated in eq. (\ref{false}) must be false, in which case we can conclude that the {\em finite-temperature} correlators are single-valued functions $\rho_{ij}^{\alpha, \beta}[\hat{W}]$ of the equilibrium density operator $\hat{W}$, which is, in turn, uniquely determined by the coupling constants $\{ J_{ij}^{\alpha, \beta} \}$ of the model, so that $\rho_{ij}^{\alpha, \beta}(J_{ij}^{\alpha, \beta})$ is injective.  

Proving the injectivity of $\rho_{ij}^{\alpha, \beta}(J_{ij}^{\alpha, \beta})$ suffices to show it is bijective (i.e., a one-to-one mapping) since $\rho_{ij}^{\alpha, \beta}(J_{ij}^{\alpha, \beta})$ is surjective by construction, assuming, of course, that the physical system under study can be described by a Hamiltonian of the form given in eq. (\ref{H}). Indeed, given a model defined by a set of coupling parameters $\{ J_{ij}^{\alpha, \beta} \}$, we can always determine, at least in principle, the respective equilibrium density operator $\hat{W}$, which can then be used to compute the finite-temperature spin-spin correlators $\{ \rho_{ij}^{\alpha, \beta} \}$ per eq. (\ref{correlator_T_eigbasis}). This is entirely analogous to the trivial surjectivity of the mapping of the ground state wave functions onto the set of number densities in the proof of the Hohenberg-Kohn theorem \cite{DreizlerGross}: every number density must be associated with a given wave function. Interestingly, the one-to-one relation between the ground state wave function and the external potential in spin-density-functional theory is not guaranteed to hold \cite{Capelle01}.

The final step amounts to recognizing that bijectivity is a sufficient condition for a function to be invertible. Hence, the coupling constants $\{ J_{ij}^{\alpha, \beta} \}$ are themselves single-valued functions of the \textit{finite-temperature} correlators $\{ \rho_{ij}^{\alpha, \beta} \}$, which concludes the generalization of Quintanilla's theorem to the case of finite temperature. Importantly, this result involves not only the ground state manifold (regardless of its degeneracy) but \textit{excited states as well}.

As in the case of Hohenberg-Kohn theorem, the present theorem does not give a systematic method to obtain the functional that relates the coupling constants $ J_{ij}^{\alpha, \beta} $ to the correlators $ \rho_{ij}^{\alpha, \beta} $. Hence, Quintanilla's theorem does not produce a practical short-term advantage to tackle quantum spin Hamiltonians. Nevertheless, this theorem does provide a solid theoretical basis for a novel approach to the important problem of determining the parent Hamiltonian of experimental systems.

Artificial intelligence methods have been used to infer spin couplings out of experimentally determined spin correlators in spin-ice compounds\cite{samarakoon2021}.  The process includes the training of an artificial neural network (ANN) based on classical spin model simulations. In a similar vein, ANNs have been trained to infer spin couplings out of specific heat measurements \cite{yu2021}. We note that our finite-temperature theorem provides not only a firm foundation but also a practical advantage to infer spin couplings out of spin correlators, as it opens the possibility of training a ANN with simulations of quantum spin models. As noted by Yu {\em et al.}\cite{yu2021}, this process is actually simplified at large temperatures. The computational resources needed to carry out exact diagonalizations of spin Hamiltonians scale exponentially with the system size. At high temperatures, however, the spatial range of spin correlators is expected to be shorter, which provides a natural cut-off for the size of the  simulation cells\cite{yu2021}. We also note that the training could be supported by digital quantum simulations of quantum spin models\cite{Kandala17} on noisy intermediate-scale quantum computers\cite{Preskill18} using hybrid variational algorithms\cite{Bharti22}. This approach may be used to accurately determine the spin Hamiltonian of Kitaev materials, such as RuCl$_3$\cite{trebst2017, Winter17}.

We finish by noting that, in all of the above, we never make use of the fact that $\hat{S}_{i}^{\alpha}$ are spin operators, and therefore the theorem  applies to any Hamiltonian that can be expressed as a bilinear sum of operators,
\begin{equation}
\hat{H} = \sum_{a,b} J_{a,b} \hat{O}_a \hat{O}_b
\label{Hgen}
\end{equation}
where $J_{a,b}$ describe couplings between operators $\hat{O}_a$ and $\hat{O}_b$, with $a$ and $b$ general labels. 
For example, Hamiltonian (\ref{Hgen})  includes the relevant case of interacting electrons that occupy a flat band or a single Landau level, with $\hat{O}$ being the electronic density operator and $J_{a,b}$ being the Coulomb interaction projected onto the lowest Landau level\cite{laughlin1983}.

The theorem stated and proven above can be rephrased as follows. The couplings $J_{a,b}$ are a single-valued functional of the thermal correlators:
 \begin{equation}
\rho_{a,b} = \frac{1}{Z}
\sum_n e^{-\beta E_n}\langle \Phi_n|
\hat{O}_a \hat{O}_b |\Phi_n\rangle 
\label{TO}
\end{equation}
Thus, the theorem establishes a one-to-one correspondence between the finite-temperature density-density correlations and the representation of the Coulomb matrix elements in the lowest Landau level. 

The Hohenberg-Kohn theorem\cite{hohenberg64}, and Mermin's finite-temperature extension\cite{mermin65}, became extremely useful when approximate versions of the density functional, such as the Kohm-Sham local density approximation\cite{KohmSham65}, were developed. We hope that this paper will inspire the quest for such approximate functionals within the context of quantum spin Hamiltonian learning based on spin structure factors. We also note the existence of a theorem relating the ground state energy to the magnetization density in Heisenberg models\cite{libero03}; its connection with our work remains to be explored.

In conclusion, we have demonstrated a theorem that establishes a one-to-one relation between interaction couplings and finite-temperature correlators in a general class of bilinear Hamiltonians. Our work generalizes a recent result of Quintanilla\cite{quintanilla22} in two ways. First, our theorem establishes the validity of Quintanilla's result for arbitrary temperatures. Second, we note that the theorem is applicable beyond the realm of spin systems. Our theorem puts the recent work that uses artificial intelligence to determine spin couplings\cite{samarakoon2021} on firm theoretical footing and may provide a route to settle disputes about the nature of spin couplings in quantum materials, such as Kitaev materials. 

{\em Acknowledgements.} J.F.R.  acknowledges financial support from  FCT (Grant No. PTDC/FIS-MAC/2045/2021),
FEDER / Junta de Andaluc\'ia --- Consejer\'ia de Transformaci\'on Econ\'omica, Industria, Conocimiento y Universidades,
(Grant No. P18-FR-4834), and Generalitat Valenciana funding Prometeo20XXX.
MICIIN-Spain (Grant No. PID2019-109539GB-C41). B.M. acknowledges support from the FCT PhD scholarship No. SFRH/BD/08444/2020.

\appendix

\section{Infinite-Temperature Spin-Spin Correlators}

The spin-spin correlators $\rho_{ij}^{\alpha \beta}(\hat{W})$ at a nonzero temperature $T$ are given by eq. (\ref{correlator_T_eigbasis}). Setting $T \to \infty$, or $\beta \to 0$, gives $e^{-\beta E_n} = 1$ for all eigenenergies $\{ E_n \}$, in which case
\begin{equation*}
    \rho_{ij}^{\alpha \beta}(T \to \infty) = \frac{1}{Z} \sum_{n} \langle \Phi_n | \hat{S}_{i}^{\alpha} \hat{S}_{j}^{\beta} | \Phi_n \rangle \equiv \frac{1}{Z} \textrm{Tr} \left( \hat{S}_{i}^{\alpha} \hat{S}_{j}^{\beta} \right).
\end{equation*}
Being a scalar, $\rho_{ij}^{\alpha \beta}(T \to \infty)$ is invariant under a change of basis. Since the density operator $\hat{W}$ now only contributes a constant prefactor $\frac{1}{Z}$, we can replace the Hamiltonian eigenbasis $\{ |\Phi_n\rangle \}$ with the product basis $\{ \bigotimes_{i} |S_i\rangle \}$, where we define the quantization axis such that $\hat{S}_{i}^{z} |S_i\rangle = S_i |S_i\rangle$ at every site $i$, with $S_i \in \{ -S, -S+1, ..., -1, 0, 1, ..., S-1, S \}$ for a local spin-$S$. Computing the trace in this product basis gives
\begin{equation*}
    \rho_{ij}^{\alpha \beta}(T \to \infty) = \frac{\delta_{\alpha z} \delta_{\beta z}}{(2S+1)^2} \sum_{S_i, S_j = -S}^{S} \langle S_i | \hat{S}_{i}^{\alpha} | S_i \rangle \langle S_j | \hat{S}_{j}^{\beta} | S_j \rangle,
\end{equation*}
where the Kronecker deltas follow from the fact that the only spin component with nonzero entries along the diagonal is $\hat{S}^{z}$, so for any choice other than $(\alpha, \beta) = (z,z)$ the correlator vanishes. The prefactor results from the fact that the trivial sums over the remaining $N-2$ spins-$S$ give a factor $(2S+1)^{N-2}$, which cancels with the partition function $Z = (2S+1)^N$.

Considering, for the moment, the nonlocal case $i \neq j$, we realize that, for every configuration where $\langle S_i | \hat{S}_{i}^{z} | S_i \rangle \langle S_j | \hat{S}_{j}^{z} | S_j \rangle$ takes a value $c$, there is another one taking the symmetric value $-c$. In other words, any term $\langle S | \hat{S}_{i}^{z} | S \rangle \langle S' | \hat{S}_{j}^{z} | S' \rangle$ is cancelled out by another term $\langle -S | \hat{S}_{i}^{z} | -S \rangle \langle S' | \hat{S}_{j}^{z} | S' \rangle$. Hence, all nonlocal spin-spin correlators vanish at infinite temperature:
\begin{equation*}
    \rho_{ij}^{\alpha \beta}(T \to \infty) = \frac{\delta_{ij} \delta_{\alpha z} \delta_{\beta z}}{2S+1} \sum_{S_i = -S}^{S} \langle S_i | (\hat{S}_{i}^{z})^2 | S_i \rangle.
\end{equation*}
The remaining sum can be computed explicitly: $\sum_{S_i = -S}^{S} S_i^2 = \frac{S(S+1)(2S+1)}{3}$. Replacing in the expression above yields
\begin{equation*}
    \rho_{ij}^{\alpha \beta}(T \to \infty) = \delta_{ij} \delta_{\alpha z} \delta_{\beta z} \frac{S(S+1)}{3}.
\end{equation*}
Of course, this result is valid for any Hamiltonian, since neither the eigenspectrum nor the eigenstates appear in any step of this calculation at infinite temperature.

\bibliography{main}{apsrev4-1}

\end{document}